\documentclass[aps,prl,twocolumn,showpacs,superscriptaddress,groupedaddress]{revtex4}

\usepackage{amsmath,amssymb,amsfonts,subfigure,braket,color,dsfont}
\usepackage[english]{babel}
\makeatletter\AtBeginDocument{\let\@elt\relax}\makeatother 
\usepackage{graphicx} 
\usepackage{sistyle}
\usepackage{listings}
\usepackage[percent]{overpic}
\usepackage{gensymb} 
\usepackage{xcolor}
\usepackage{tikz} 
\usepackage{flushend}
\usepackage{ulem}

\begin{document} 
 
\title{From Heat Capacity to Coherence in Ultra-Narrow-Linewidth Solid-State Optical Emitters at Sub-Kelvin Temperatures}


\author{D. Serrano} 
\affiliation{Chimie ParisTech, Universit\' e PSL, CNRS, Institut de Recherche de Chimie Paris, 75005 Paris, France} 
\author{T. Klein} 
\affiliation{Univ. Grenoble Alpes, CNRS, Grenoble INP and Institut N\' eel, 38000 Grenoble, France}
\author{C. Marcenat} 
\affiliation{Univ. Grenoble Alpes, CEA, Grenoble INP, IRIG, Pheliqs, 38000, Grenoble, France}
\author{P. Goldner} 
\affiliation{Chimie ParisTech, Universit\' e PSL, CNRS, Institut de Recherche de Chimie Paris, 75005 Paris, France} 
\author{M. T. Hartman} 
\altaffiliation{Present address: Institut Fresnel, UMR 7249, Université d'Aix-Marseille, CNRS, Centrale
Méditerranée, 13013 Marseille, France.}
\affiliation{Laboratoire Temps Espace (LNE-OP), Observatoire de Paris, Université PSL, Sorbonne Université, Université de Lille, LNE, CNRS, 75014 Paris, France}
\author{B. Fang} 
\affiliation{Laboratoire Temps Espace (LNE-OP), Observatoire de Paris, Université PSL, Sorbonne Université, Université de Lille, LNE, CNRS, 75014 Paris, France}
\author{Y. Le Coq}
\affiliation{Univ. Grenoble Alpes, CNRS, LIPhy, 38000 Grenoble, France}
\author{S. Seidelin}\email{signe.seidelin@neel.cnrs.fr}
\affiliation{Univ. Grenoble Alpes, CNRS, Grenoble INP and Institut N\' eel, 38000 Grenoble, France}

\date{\today}  

\begin{abstract}

The coherence properties of optical emitters in crystals are crucial for quantum technologies and optical frequency metrology. Cooling to sub-kelvin temperatures can markedly enhance coherence, making it important to identify the parameters governing emitter and host crystal behavior in this regime. We investigate a Czochralski-grown europium-doped yttrium orthosilicate crystal, reporting measurements of its heat capacity and optical coherence. Heat capacity not only informs thermal noise limits in metrology schemes but can also reveal two-level systems (TLS) arising from crystal imperfections via a linear-in-temperature term. Below 1 K, where phonon contributions are suppressed, TLS can drive decoherence, leading to a linear broadening of the homogeneous linewidth. From our data, we place an upper bound on the TLS contribution. This, together with constant optical linewidths between 300 mK and 2 K measured via photon-echo lifetimes, is consistent with a minimal TLS effects in our sample. A low level of TLS is particularly important for the performance of optical quantum devices based on doped crystals, since their presence could otherwise limit further improvements in coherence at sub-kelvin temperatures.

\end{abstract}

\pacs{65.40.-b,42.50.Ct.,76.30.Kg}

\maketitle

\section{\label{intro} Introduction}

As scalable quantum technologies progress, demands on the underlying physical platforms continue to grow. Rare-earth ions embedded in crystals are known for exceptional optical coherence, even at liquid helium temperatures, achieving record-narrow linewidths among solid-state optical emitters. For instance, spectral hole burning (SHB) can produce structures with linewidths below a few kHz, offering a viable alternative to ultrastable cavities in optical frequency metrology~\cite{Julsgaard2007,Thorpe2011,Galland2020_OL,Lin2023_OE,Gustavsson2025}. Rare-earth doped crystals are also used in quantum memories~\cite{Zhong2017,Laplane2017,Chao2020}, photon interfaces~\cite{Rakonjac2021,Rivera2023}, and hybridized with solid-state nuclear spins~\cite{OSullivan2024} or superconducting qubits~\cite{Gouzien2021}, and have also emerged as a promising platform for optomechanics~\cite{Molmer2016,Seidelin2019, Otha2021,Chauvet2023}.

As these systems have been increasingly optimized and other broadening mechanisms suppressed, temperature-dependent $T^7$ broadening from two-phonon Raman scattering~\cite{Konz2003} can emerge as a limiting factor, motivating operation below the 1 K regime. Recent measurements on $\rm Eu^{3+}$-doped $\rm Y_2SiO_5$ (Eu:YSO) crystals have shown that operating at 300 mK suppresses frequency fluctuations due to thermal instability~\cite{Lin2024_PRL}. However, despite improved frequency stabilitywe recurrently observe a weak linear broadening of spectral holes with temperature in our crystal—on the order of 1 kHz/K—even where $T^7$ broadening is negligible~\cite{Lin2025}. A similar behavior has been observed in silicon-vacancy centers in diamond~\cite{Becker2017}, and more pronounced linear broadening has been reported between 1.5 and 5.5 K in photon-echo measurements of europium ions in other hosts~\cite{Schmidt1994,Flinn1993,Macfarlane2004,Kunkel2016}, as well as in other rare-earth ions~\cite{Macfarlane2000}. These effects have consistently been attributed to disorder modes or two-level systems (TLS)~ \cite{Phillips1972,Anderson1972}, typically associated with amorphous materials or crystals with some degree of disorder. 
Therefore, although our crystal was grown using an optimized Czochralski technique to ensure high structural quality, it may still contain a sufficiently high TLS density to account for the observed broadening. In this work, we quantitatively investigate this possibility using complementary methods applied to the same crystal that exhibited linear broadening in the SHB experiments. Indeed, even minute imperfections can become significant at sub-kelvin temperatures, motivating a quantitative assessment of TLS in our Eu:YSO system. Heat capacity measurements provide a known diagnostic, as TLS manifest through a linear-in-temperature term. More precisely, in the standard tunneling model of two-level systems in disordered solids, the density of states is assumed to be constant at low energies, leading to a heat-capacity contribution linear in temperature~\cite{Phillips1972,Anderson1972,Enss2005}. In addition, we compare these results with new measurements of the temperature-dependent linewidth obtained using a different technique: while previous sub-kelvin studies have relied solely on spectral hole burning~\cite{Lin2024_PRL}, we implement 2-pulse photon-echo measurements to probe line broadening in this novel temperature regime.

\begin{figure}[t]   
    \includegraphics[width=7cm]{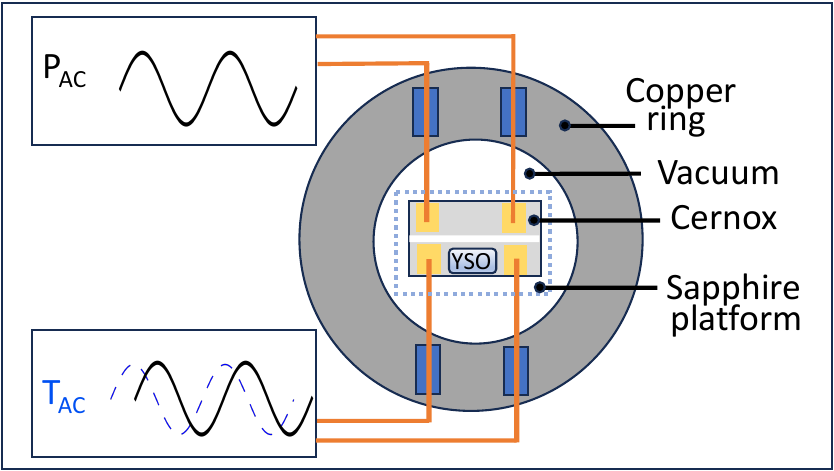}
    \caption{\label{exp} Schematic of the setup used for heat-capacity measurements. The europium-doped YSO sample is mounted on a Cernox chip (light grey rectangle), which is cleaved into two halves (separated by the white line). The suspended chip is connected (yellow pads) by electrical wires to a connection block (blue pads) that is in thermal contact with a copper ring, itself screwed onto the cold finger of the cryostat. While this contact provides good thermal anchoring, it remains electrically isolated. Separate wiring is used to provide electrical connections for the heating current and for resistance-measurement electrical currents. An oscillatory heating power $P_{AC}$ is applied to the crystal through the first half of the Cernox, inducing temperature oscillations $T_{AC}$ measured in the second half. The amplitude and phase of $T_{AC}$ are measured to extract the crystal's heat capacity.}
\end{figure} 

\section{\label{System} The $\rm Eu^{3+}$:$\rm Y_2SiO_5$ system}

All measurements are conducted on bulk $\rm Eu^{3+}$ doped (0.1 atomic \%) $\rm Y_2SiO_5$ crystals fabricated in-house. The crystal was grown from the melt using the Czochralski method using an inductively heated iridium crucible. The starting oxides were of at least 99.99 \% purity. The crystal was pulled from the melt along the [010] direction. The resulting crystal boule was 1 inch in diameter, colorless, and free from cracks. The boule was subsequently oriented using Laue x-ray diffraction, and cut perpendicular to the $b$-axis with a precision better than 1 degree \cite{Ferrier2016}. 

Depending on the experiments, the samples are then prepared in two different ways. They all originate from the same crystal boule as the one that earlier exhibited linear broadening effects. For the low-temperature heat capacity measurements, a small piece of the crystal is cut into a sub-mm$^3$ piece, with a total mass of 1.89 mg. The YSO density being 4.45 g/cm$^3$ this corresponds to a crystal volume of 0.42 mm$^3$. For the optical measurements, the crystal is cut along the crystallographic $b$-axis and the dielectric D1 and D2 axes into a 8 mm $\times$ 8 mm $\times$ 4 mm cuboid, with the largest facets perpendicular to the $b$-axis and polished to an optical finish.

The europium ions can substitute yttrium atoms in two distinct, nonequivalent sites within the YSO matrix, known as site 1 and site 2 (vacuum wavelengths of 580.04\,nm and 580.21\,nm, respectively \cite{Thorpe2011}). These wavelengths are associated with the optical transition $^7F_0 \rightarrow$  $^5D_0$~\cite{Equall1994,Oswald2018}. In the 2-pulse photon-echo experiments discussed below, we focus on ions occupying site 1.

\section{\label{heat_capacity}Measurements of the heat capacity}

In laser stabilization using optical cavities, the heat capacity of the cavity components (mirror substrate and coatings as well as the spacer material) is a critical parameter for estimating thermal-noise-limited frequency instability using the Fluctuation Dissipation Theorem. It influences the system’s response to temperature fluctuations, Brownian motion, thermo-optic effects, thermal expansion, and energy dissipation - all of which degrade cavity stability \cite{Levin1998,Braginsky1999,Numata2004}. Similarly, in alternative approaches where the laser is stabilized to a spectrally burned hole, such as in Eu:YSO, the current state-of-art, the heat capacity of the crystal remains a key factor in determining the frequency stability limited by thermal noise \cite{Hartman2024}. Finally, as discussed, through the presence of a linear term, heat capacity can provide insight into the presence of TLS that might impose a fundamental limit on emitter coherence. In the following, we present our measurements of heat capacity to better understand these effects and to set an upper bound on their impact on coherence properties, which are the subject of the second part of the article.
 
\begin{figure}[t]
    \centering
    \begin{tikzpicture}
        \node[anchor=south west,inner sep=0] (image) at (0,0) {\includegraphics[width=0.45\textwidth]{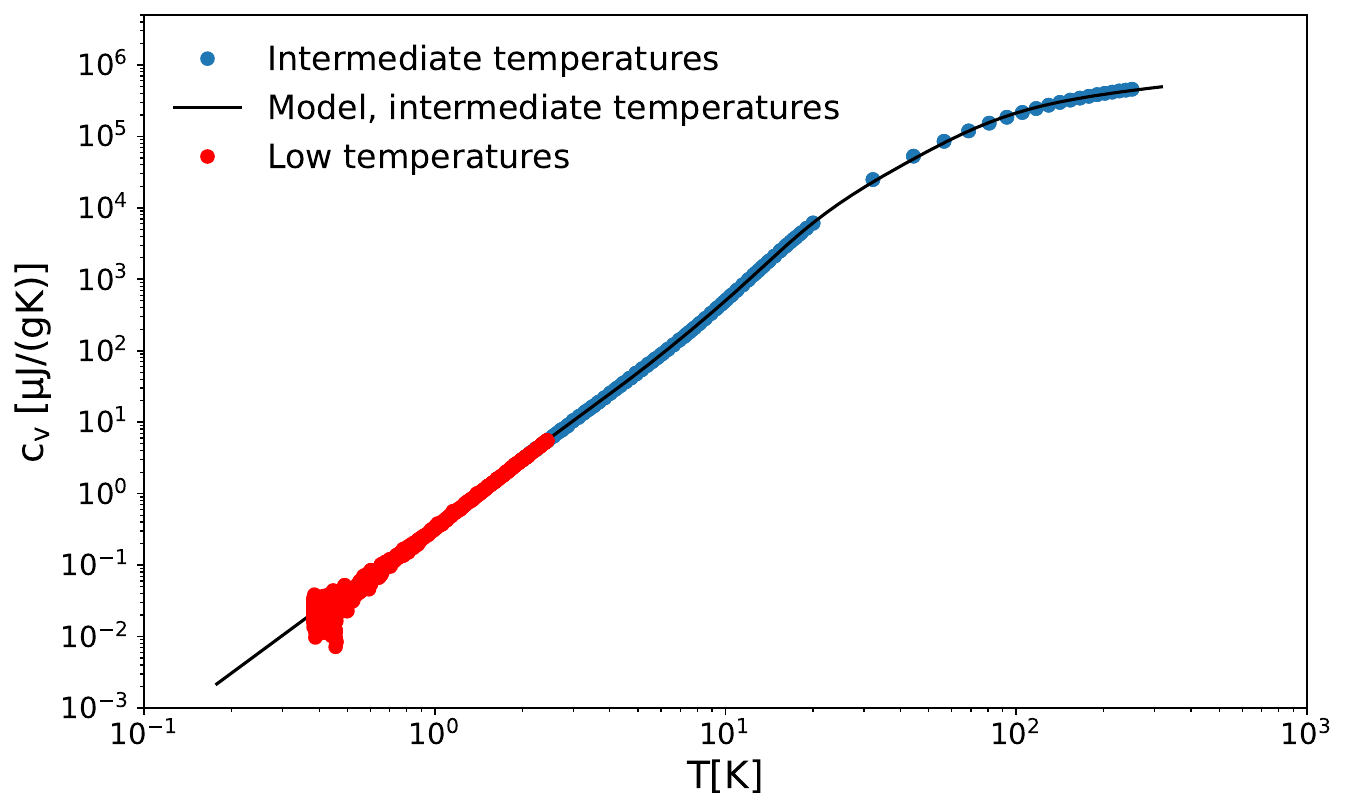}};
        \begin{scope}[x={(image.south east)},y={(image.north west)}]
            \node[anchor=south west] at (0.51,0.14) {\includegraphics[width=0.19\textwidth]{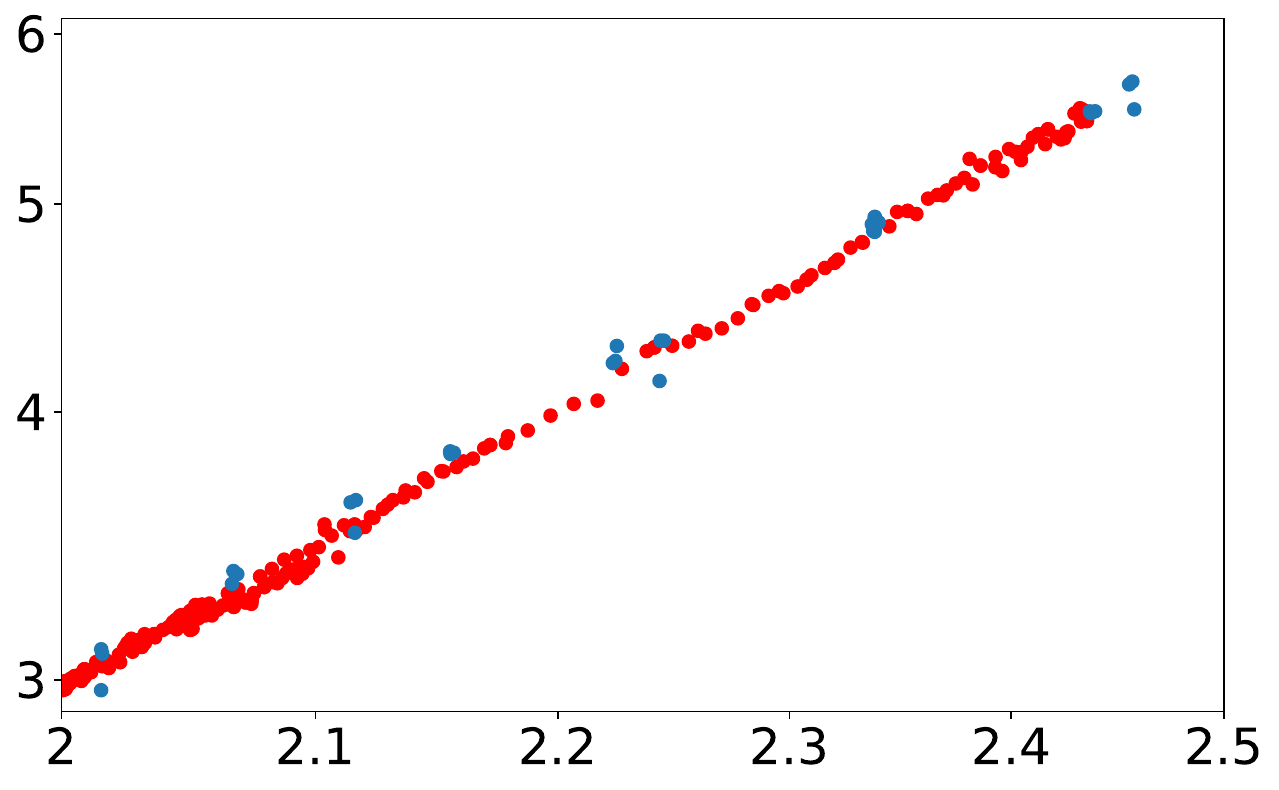}};
        \end{scope}
    \end{tikzpicture}
    \caption{Specific heat capacity measurements for the 0.1 \% Eu:YSO system across both ultralow and intermediate temperature ranges. The data in the intermediate region are modeled using a Debye integral model. The inset (same units as in main graph) highlights the region of overlap between the two datasets.}
    \label{SU_and_NEEL}
\end{figure}

\begin{figure*}[t]
    \centering
    \begin{minipage}[b]{0.41\textwidth}
        \centering
        \includegraphics[width=\textwidth]{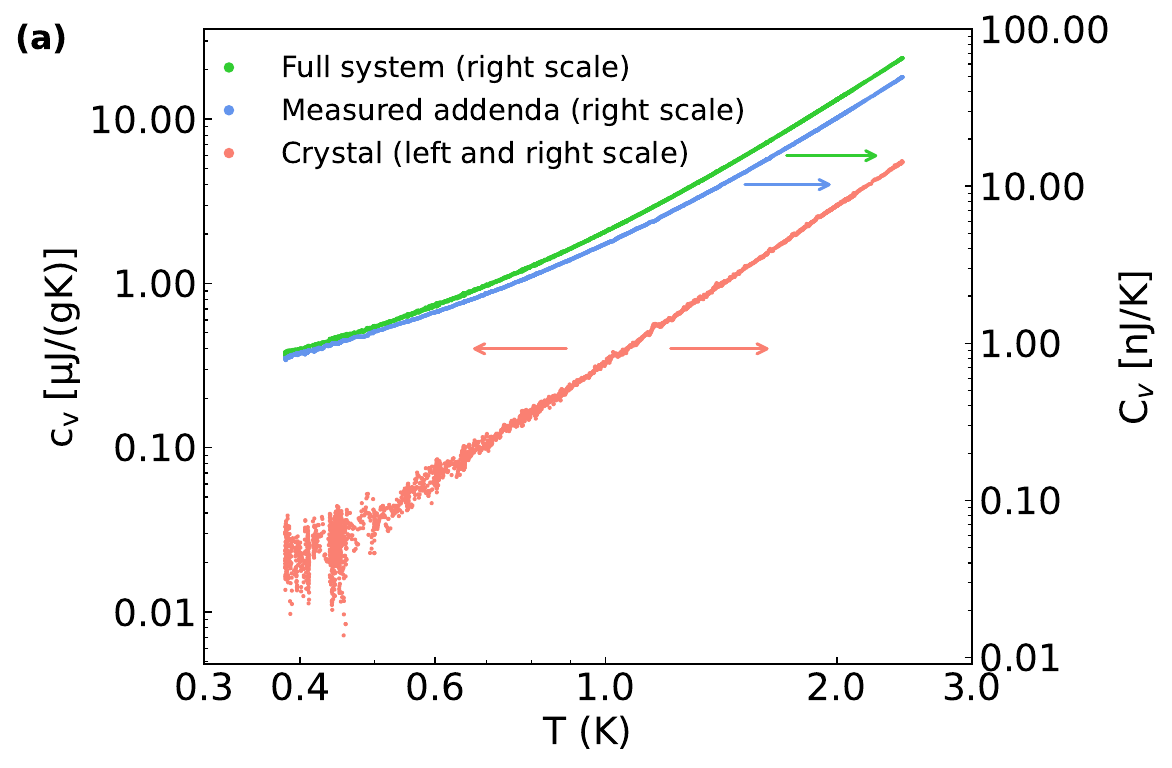}
    \end{minipage}
    \hspace{0.2cm}
    \begin{minipage}[b]{0.41\textwidth}
        \centering
        \includegraphics[width=\textwidth]{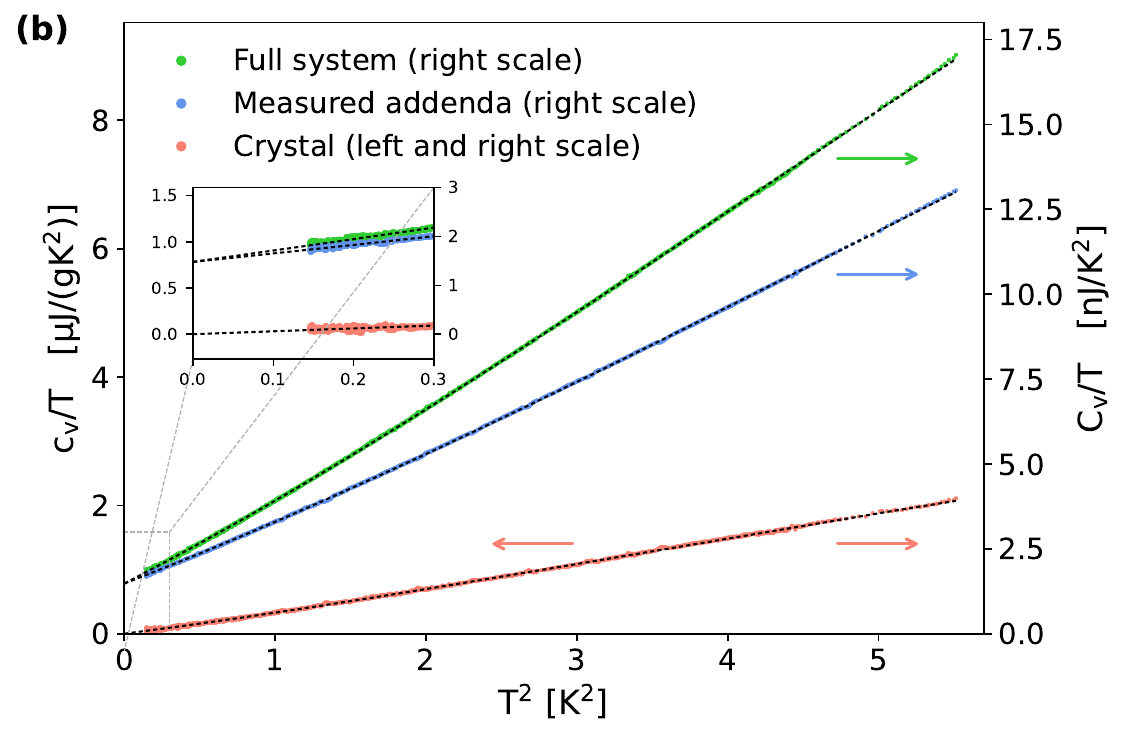}
    \end{minipage}
    \caption{Heat capacity measurements $C_v$ of the 0.1\% Eu:YSO system measured in the low temperature range:
        Full system (green), independently measured addenda (blue), and crystal alone (red), obtained by subtracting the addenda and an additional grease contribution from the full system, all read out on the right-hand $y$-axis.
        The specific heat capacity \( c_v \) for the crystal alone, obtained by dividing the \(C_v\) with the crystal mass, is shown on the left-hand \( y \)-axis (thus valid only for the red curve; also note the scale change from nJ to \( \mu\mathrm{J} \)).
        (a) Double logarithmic plot and (b) linear plot of the same data but divided by temperature and plotted as a function of temperature squared, with their associated fits (dotted black lines) such that the linear-in-temperature component appears as the intercept with the \( y \)-axis, the inset (same units as in main graph) magnifying the relevant region.
    }
    \label{heat}
\end{figure*}  
    
The heat capacities at low temperatures (380 mK to 2.4 K) are measured using a 1.89 mg crystal in a $^3$He refrigerator cryostat using a high-sensitivity
lock-in modulation technique, developed in Refs. \cite{Marcenat2015,Kacmarcik2018,Michon2019} and illustrated in Fig.~\ref{exp}. Conceptually, a periodically modulated heating power, $P_{AC}$, is applied to the crystal at a frequency $\omega$, inducing temperature oscillations at the same frequency. The amplitude  of these oscillations, $|T_{AC}|$, and its thermal phase shift $\phi$ relative to the applied power are obtained using lock-in amplifier demodulation techniques. The total heat capacity is then calculated using the following formula: $C_v= \frac{|P_{AC}| \sin(-\phi)}{\omega |T_{AC}|}$. The working frequency (here $\omega=2 \pi\times$ 2 Hz) is chosen carefully after testing the thermal homogeneity of the thermometer with the sample such that the phase is kept in the range where $-60 \degree < \phi < -30 \degree$. Experimentally, the sample is glued with a minute amount of Apiezon N grease onto the bare back surface of a miniature Cernox chip, which is cleaved into two parts, one acting as a heater (by applying an electrical current), and the other acting as a temperature sensor after calibration against a ruthenium oxide thermometer. The front surface of the chip is bonded to a sapphire substrate to ensure thermal contact between the two parts. This chip is then mounted on a small copper ring using platinum–tungsten wires (7 \% tungsten), chosen because their intrinsic thermal conductivity allows access to the desired range of phase shifts while minimizing the wires’ contribution to the total heat capacity of the system. This set-up enables us to obtain absolute values of the specific heat with an reproductibility significantly better than 95\% as deduced from measurements on ultra-pure copper \cite{Michon2019}. 

After measuring the total heat capacity of the system, the heat capacity of the crystal alone is obtained by subtracting the contributions from the measurement addenda, which include a sapphire platform, a small amount of silver epoxy used for electrical contacts, the connecting wires, and the small amount of grease used to affix the crystal to the platform. To enable this subtraction, an independent measurement of the addenda alone is performed separately. However, slight variations in the amount of grease between the full-system and addenda-only measurements can lead to a systematic error in the extracted crystal heat capacity. To correct for this potential error and provide an independent consistency check, we perform an independent specific heat measurements over an extended temperature range (2–250 K) on a significantly larger crystal (31 mg), again from the same crystal boule. These measurements were carried out using a standard thermal relaxation method with a Physical Property Measurement System (Quantum Design) equipped with the “Heat Capacity” option. By overlapping the data-sets in the common temperature interval (2–2.4 K), we are able to quantify the grease contribution omitted in the addenda measurement in the low temperature range. This analysis yields an unaccounted grease mass of (21.3 $\pm$ 0.3) $\mu$g. Using highly precise measurements of the heat capacity of this type of grease in the relevant temperature range from the literature \cite{Schink1981,Schnelle1999}, we correct our original data set accordingly. The results for the specific heat capacity of the crystal are presented in Fig.~\ref{SU_and_NEEL}. The main panel displays both the low- and intermediate-temperature data, obtained using different experimental setups at separate institutions. The intermediate-temperature data are accompanied by a model appropriate for this regime, which combines the Debye integral approach with two additional Einstein modes, following the methodology described in Ref.~\cite{Gamsjager2018}. The inset zooms on the data overlap region, enabling the precise determination and correction of the grease mass.

In the following we focus on the low temperature data, presented in Fig.~\ref{heat}. To obtain the heat capacity for the crystal alone, we subtract the contribution of the addenda, measured separately under identical conditions.  Well below the Debye temperature, approximately 500 K for $\rm Y_2SiO_5$~\cite{Denault2015} the heat capacity can be written as a polynomial of $T$:

\begin{equation}\label{eq1}
c_v=\alpha T + \beta T^3 + \gamma T^5 + \delta T^7.
\end{equation}

Here, the $T^3$ term corresponds to the contribution of acoustic phonons in the Debye approximation, while the higher-order terms arise from nonlinearities in the phonon dispersion. At very low temperatures ($T \leq$ 1\,K), a term proportional to $T$ may appear if two-level systems (TLS) are present in the material. The absence of even order terms is a direct consequence of the bosonic phonon population integrals and symmetry properties of the expansion, see for instance reference~\cite{Collings1986} for details. We have truncated the expansion in Eq.~\ref{eq1} at $T^{7}$ based on the Akaike Information Criterion~\cite{Akaike1974}, which shows a clear improvement in the fit quality up to this order, but only a marginal gain beyond, insufficient to justify the increased model complexity and the risk of over-parameterisation.
 
In fig.~\ref{heat} (a) we have shown the data corresponding to the heat capacity for the full system, the addenda and the crystal, as well as the specific heat capacity of the crystal. Figure~\ref{heat} (b) plots the same quantities but divided by temperature, and as a function of $T^2$. The fit function therefore takes the form of $c_v/T= \alpha + \beta T^2 + \gamma T^4 + \delta T^6$ which allows the linear contribution to manifest visually as the vertical intercept $\alpha$ upon extrapolation to zero temperature. The linear term $\alpha$ is then, according to the discussion above, associated with the contribution from TLS to the heat capacity: $c_v^{\rm TLS}=\alpha T$ ~\cite{Phillips1972,Anderson1972,Enss2005}.  As the data is plotted as a function of $T^2 \equiv x$, the fit function takes the simple cubic form $\alpha + \beta x + \gamma x^2 + \delta x^3 $, and we obtain the following coefficients and their associated uncertainties: $\alpha = (0.0~\pm~ 2.5)~\rm nJ/(g K^2)$, $\rm \beta= (306~ \pm~ 5)~ nJ/(g K^4)$, $\rm \gamma= (25~ \pm~ 4)~ nJ/(g K^6)$ and  $\rm \delta= (-2.2~\pm~ 1.0)~nJ/(g K^8),$ as expected largely dominated by the Debye phonon-term proportional to $\beta$. The uncertainties include statistical errors and variations due to fitting range. Thus, we measure $\alpha$ to be 0 within the uncertainty.

We note that, at a Eu concentration of only 0.1\%, the measured specific heat capacity essentially reflects that of the YSO host crystal, largely independent of the specific dopant ion. Accordingly, these results should be applicable to YSO crystals with any low ion-dopant concentration. 

As to our knowledge, specific heat capacity measurements have not priorly been performed in YSO at these temperatures, however, we can compare with a measurement obtained in europium-doped silicate glass \cite{Schmidt1994}, a system expected to exhibit a much higher degree of TLS due to the absence of a crystalline structure. In this work, the authors determined a value of $ c_{v}^{\rm TLS}  = 3.5 \times T~ [\micro$J/(g K)], making our upper limit,  $ c_{v}^{\rm TLS}  = 2.5 \times T~ [n$J/(g K)], a factor of 1000 lower, indicating that any TLS contribution in YSO is correspondingly weaker and therefore intrinsically more challenging to resolve experimentally.
    
\begin{figure*}[t]
    \centering
    \begin{minipage}[b]{0.4\textwidth}
    \centering
    \begin{overpic}[width=\textwidth]{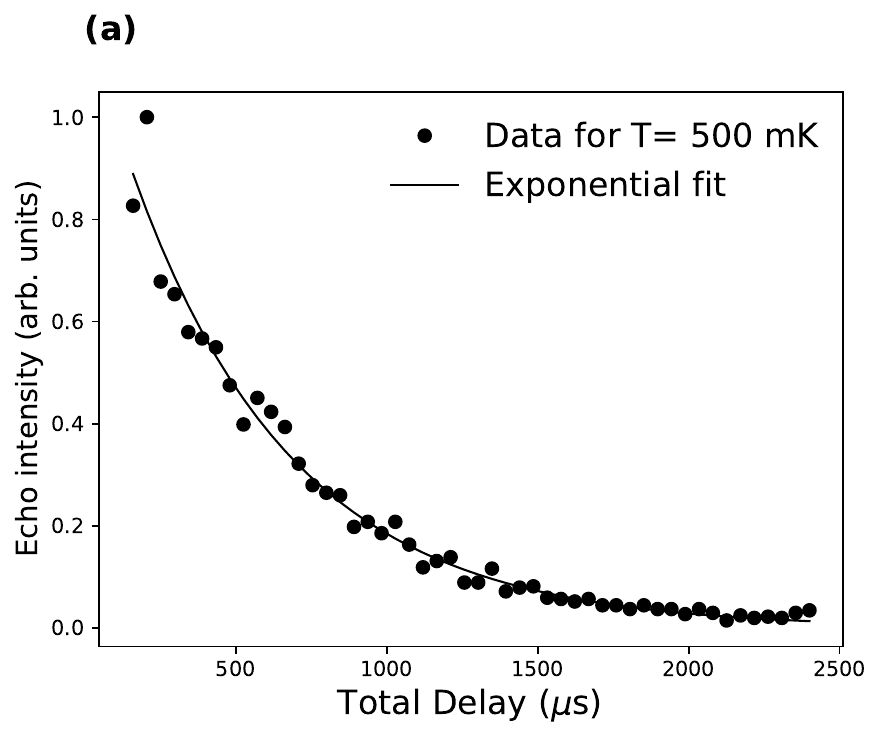}
        \put(30,79){\textbf{$\rm  T_2=(535 \pm 17) ~\mu s$}}
    \end{overpic}
    \end{minipage}
  \hspace{0.6 cm}
    \begin{minipage}[b]{0.45\textwidth}
        \centering
        \begin{overpic}[width=\textwidth]{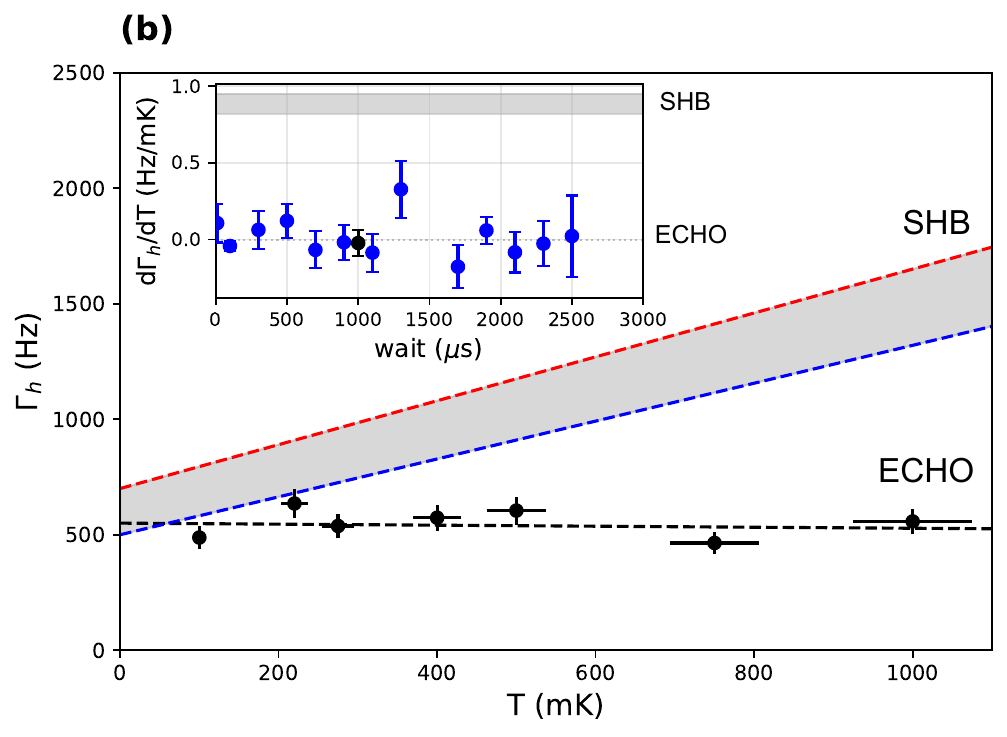}
        \put(30,70){$\rm \Gamma_{h,av}=(573 \pm 35)~ Hz$}
    \end{overpic}      
    \end{minipage}
    \caption{Coherence times and linewidths derived from photon-echo measurements. Panel (a) shows a representative 2-pulse photon-echo intensity trace recorded at 500 mK, corresponding to a homogeneous linewidth $\rm \Gamma_h=1/(\pi T_2)=(595 \pm 19)~ Hz$. In panel (b), homogeneous linewidths measured by 2-pulse photon-echo as a function of temperature (black filled circles) with the associated average linewidth $\rm \Gamma_{h,av}$. The two colored dashed lines (red and blue) indicate the results obtained in former Spectral Hole Burning (SHB) experiments \cite{Lin2025}. The black dashed line is a linear fit of the data.  The blue data points in the inset in (b) correspond to the slope for 3-pulse photon-echo measurements, for variable wait times for the sequence, whereas the grey zone indicates the SHB measurement for comparison (in this case, obtained on a timescale of 2-3 seconds). The black dot corresponds to the 2-pulse result shown in the main figure.}
   \label{echo}
\end{figure*} 



\section{Optical linewidth broadening and photon-echo measurements}

As discussed above, a presence of TLS has been associated with a linear broadening of the homogeneous linewidth in solid state emitters. In the europium-doped silicate glass system cited above, based on spectral hole burning within the inhomogeneously broadened europium absorption line, a broadening of the linewidth of approximately 9 MHz/K was observed below 4 K, indicating the presence of TLS~\cite{Schmidt1994}. As mentioned, in our crystal, a much weaker linear temperature dependence of the spectral hole linewidth ($\sim$ 1 kHz/K) is observed in the sub-1 K regime~\cite{Lin2025}, consistent with a significantly lower TLS density. Assuming that the coupling strength between TLS and europium ions is similar in silicate glass and our YSO crystal, scaling our $\alpha$ by the difference in broadening observed would lead to $\alpha = 0.35$ nJ/(gK$^2$), compatible with our conservative upper limit, and even consistent with the smaller uncertainty arising from the statistical uncertainty alone (0.4 nJ/(gK$^2$)). Note that the assumption of similar coupling strengths is not obvious; here, we have used it solely to provide a consistency check between the orders of magnitude observed across the different systems. 

It is also worth noting that the influence of TLS on the observed linewidth may depend on the specific measurement technique used to determine the broadening, due to the involvement of different characteristic timescales. In both our former experiments~\cite{Lin2025} and those conducted on silicate glass cited above~\cite{Schmidt1994}, linewidth was assessed via the spectral hole burning requiring a measurement sequence of the order of   2-3 seconds, introducing a delay between the burning of the hole and finalizing the measurement of the linewidth by scanning the laser across the profile. An alternative approach is based on photon echo experiments, which typically explore the linewidth established over a milisecond timescale. In the following, we examine whether the influence of TLS on linewidth is still observable on this short timescale. The optical setup is similar to the one described in Ref.~\cite{Ferrier2016} but here performed in a Bluefors DS dilution refrigerator cryostat. Photon-echo experiments are performed using 6 $\mu s$-long laser pulses focused to a spot diameter on the order of 1 mm on the crystal. The laser wavelength is set to 580.04217 nm, with a power of 10 mW—sufficiently low to avoid excitation-induced decoherence~\cite{Thiel2014}. The temperature is varied from 100 mK to 1 K in 100 mK increments, and at each temperature, the photon-echo intensity recorded. Pulse shaping of the excitation light is achieved using an acousto-optic modulator in a double-pass configuration. To minimize persistent spectral hole burning effects, the laser frequency is continuously scanned over a 300 MHz range within a 2 s period.  We have performed both 2- and 3-pulse (varying wait times between second and third pulse) photon-echo measurements, the results being plotted in Fig.~\ref{echo}. Panel (a) presents the 2-pulse photon-echo decay measured at 500 mK, from which the homogeneous linewidth is determined, while the main plot in panel (b) displays 2-pulse photon-echo linewidths measured at various temperatures. In addition, the data points in the inset correspond to the slope $d\Gamma_h/dT$ for 3-pulse photon-echo measurements for variable wait times. The errors in the linewidths $\Gamma_h$ corresponding to the statistical uncertainties extracted from the exponential fits are smaller than the size of the data marker. The dominant uncertainty, as indicated by the vertical errorbars in (b), arises from experiment repeatability, in particular cryostat vibrations and laser instabilities.


Interestingly, over the investigated temperature range (100 mK to 1 K), the photon-echo measurements reveal no measurable temperature dependence of the linewidth within the experimental uncertainty, contrary to the SHB experiments although performed with the same crystal. More specifically, the homogeneous linewidth extracted from 2-pulse photon-echo measurements remains essentially constant throughout the explored temperature interval. The same conclusion holds for the 3-pulse photon-echo measurements, performed with waiting times of up to 2.5 ms between the second and third pulse. As evidenced by the slope, d$\Gamma_h$/dT, which is compatible with zero within error bars, no significant temperature-induced broadening is observed.


The observed discrepancy in Fig.~\ref{echo} (b) between the temperature dependence of coherence times from photon echo measurements and earlier reports of linewidths from SHB measurement may, as suggested, arise from the different timescales probed by the two techniques. Decoherence due to spectral diffusion driven by phonon-mediated fluctuations of the local strain field in the host crystal has been associated with TLS or strain-mediated ion–ion interactions \cite{Louchet-Chauvet2023}. In high mechanical quality factor monocrystals such as ours, this relaxation can exceed tens of milliseconds \cite{Wagner2024}, potentially explaining its absence in photon echo measurements. The commonly assumed linear relation between TLS density—proportional to $c_{v}^{\rm TLS}$—and linewidth presumes a measurement timescale longer than the TLS tunneling time \cite{Anderson1972}. Such timescale dependence has been reported in rare-earth-doped ceramics \cite{Kunkel2016}, and may account for the differing temperature behaviors observed in photon echo and spectral hole burning.  In contrast to the ceramics, our crystal system —owing to the narrow linewidth of europium ions in YSO— requires a laser with higher stability to reach similar delay ranges as those in Ref. \cite{Kunkel2016}, necessary to probe intermediate-timescale dynamics.

\section{\label{Conclusion} Conclusion}

Our investigation into the thermal and optical coherence properties of Eu:YSO at sub-Kelvin temperatures has yielded insights into parameters essential for frequency metrology and quantum technologies.  In particular, we have measured the heat capacity and shown that, in the low-temperature range studied here (380 mK–2.4 K), it is predominantly governed by the phonon term proportional to $T^3$. We have also examined the linear contribution associated with two-level systems (TLS) and found it to be compatible with zero, albeit with relatively large uncertainty despite the use of advanced calorimetric techniques. Consequently, our earlier observations of linewidth broadening on the same crystal remain consistent with the presence of a linear contribution within the range permitted by the measurement uncertainty.

Complementary 2 and 3-pulse photon-echo measurements of the coherence properties at sub-Kelvin temperatures reveal no short-timescale linewidth broadening in the same crystal where spectral hole burning had previously indicated a linear temperature dependence. We attribute this discrepancy to the distinct temporal regimes probed by the two techniques, since some broadening mechanisms—potentially those associated with TLS—may not manifest on the short timescales accessible in standard photon-echo experiments. Future work will focus on extending the photon-echo measurements using an ultrastable laser to reach hundreds of milliseconds, where slower spectral-diffusion processes may become observable, as well as exploring crystals with varying TLS densities, including higher-density samples. Identifying the underlying mechanisms responsible for low-temperature broadening on longer time scales will not only clarify the role of TLS and other relaxation channels but may also enable strategies to suppress these effects and thereby enhance the coherence properties of rare-earth–doped crystals for quantum and metrological applications.
  
\section{Acknowledgements} 

SS thanks  Henrik M. Rønnow and Thierry Chanelière for useful discussions. We also thank Jérôme Debray for assistance in crystal cutting for the heat capacity measurements, Alban Ferrier for help preparing the Eu:YSO samples and David Hrabovsky for carrying out the measurements of the heat capacity between 2 K and 250 K. The authours acknowledge financial support from Ville de Paris Emergence Program, the Région Ile de France DIM SIRTEQ, the LABEX Cluster of Excellence FIRST-TF (ANR-10-LABX-48-01) within the Program ``Investissement d’Avenir'' operated by the French National Research Agency (ANR), the 15SIB03 OC18 and 20FUN08 NEXTLASERS projects from the EMPIR program co-financed by the Participating States and from the European Union’s Horizon 2020 research and innovation program, and the UltraStabLaserViaSHB (GAP-101068547) from the Marie Skłodowska Curie Actions (HORIZON-TMA-MSCA-PF-EF) from the European Commission Horizon Europe Framework Programme (HORIZON). We finally acknowledge funding from ANR via the grant ANR-24-CE47-1190. \\

\section{Data availability statement}

The data presented in this article are available from the authors upon reasonable request.


\end{document}